\documentclass[conference]{IEEEtran}

\usepackage{epsfig}
\usepackage{mathrsfs}
\usepackage{dsfont}
\usepackage{amsmath}
\usepackage{amssymb}
\usepackage{amscd}
\usepackage{latexsym}
\usepackage{array}
\usepackage{epsf}
\usepackage{cite}
\usepackage{psfrag}
\usepackage{color}
\usepackage{multirow}
\usepackage{tabularx}
\usepackage{graphicx}
\usepackage{float}
\usepackage{subfig}

\newcommand{\rect}{\mathop{\mathrm{rect}}\nolimits}

\makeatletter
\newcommand{\rmnum}[1]{\romannumeral #1}
\newcommand{\Rmnum}[1]{\expandafter\@slowromancap\romannumeral #1@}
\makeatother

\begin{document}
%
\title{Designing Power-Efficient Modulation Formats for Noncoherent Optical Systems}

\author{\IEEEauthorblockN{Johnny Karout\IEEEauthorrefmark{1},
Erik Agrell\IEEEauthorrefmark{1},
Krzysztof Szczerba\IEEEauthorrefmark{2} and
Magnus Karlsson\IEEEauthorrefmark{2}}
\IEEEauthorblockA{\IEEEauthorrefmark{1}Department of Signals and Systems\\
}
\IEEEauthorblockA{\IEEEauthorrefmark{2}Department of Microtechnology and Nanocience\\Chalmers University of Technology, SE-412 96 G\"{o}teborg, Sweden\\ Email: johnny.karout@chalmers.se}}

%

\maketitle

\begin{abstract}

We optimize modulation formats for the additive white Gaussian noise channel with a nonnegative input constraint, also known as the intensity-modulated direct detection channel, with and without confining them to a lattice structure. Our optimization criteria are the average electrical and optical power.
The nonnegativity input signal constraint is translated into a conical constraint in signal space, and modulation formats are designed by sphere packing inside this cone. Some remarkably dense packings are found, which yield more power-efficient modulation formats than previously known.
For example, at a spectral efficiency of 1 bit/s/Hz, the obtained modulation format offers a 0.86 dB average electrical power gain and 0.43 dB average optical power gain over the previously best known modulation formats to achieve a symbol error rate of $10^{-6}$. This modulation turns out to have a lattice-based structure.
At a spectral efficiency of 3/2 bits/s/Hz and to achieve a symbol error rate of $10^{-6}$, the modulation format obtained for optimizing the average electrical power offers a 0.58 dB average electrical power gain over the best lattice-based modulation and 2.55 dB gain over the best previously known format.
However, the modulation format optimized for average optical power offers a 0.46 dB average optical power gain over the best lattice-based modulation and 1.35 dB gain over the best previously known format.

%

\end{abstract}
%

%
\IEEEpeerreviewmaketitle

\section{Introduction}
%
%
%
%


\IEEEPARstart{M}{ultilevel} modulation has attracted significant research interest with its ability to improve the spectral efficiency of communication systems. The enabling technology behind it is the coherent transmission and detection, which gives access to both the carrier amplitude and phase to carry information. However, this increased spectral efficiency comes at the expense of a reduced power efficiency, which is undesirable in systems where power consumption is a constraint.
Therefore, designing modulation formats which offer a good trade-off between spectral and power efficiency becomes challenging.
Using \emph{lattice codes}, which are a finite set of points selected out of an $N$-dimensional lattice, is one approach which has been extensively used in the construction of multilevel modulation formats for additive white Gaussian noise (AWGN) channels with coherent detection~\cite{Forney1988,Forney1989,Conway1999}. In addition, techniques such as constellation shaping and nonequiprobable signaling have been used to minimize the average power\cite{Forney1989,Calderbank1990}. The former is done by selecting the set of points in a lattice which have minimum energies, whereas the latter minimizes the average power by reducing the transmission frequency of points with high energies.
Another approach is by resorting to \emph{numerical optimization} techniques to find the best possible packing of constellation points as in~\cite{Foschini1974,Porath2003,Sloane1995,Graham1990,Agrell2009} for different power constraints, whether average or peak power. The drawback is the lack of geometric regularity, which increases the modulator and demodulator complexity. As the number of constellation points increases, the best known packings approach a regular structure such as a lattice~\cite[Ch. 1]{Conway1999}.

Both approaches for designing power-efficient modulation formats assume that both the amplitude and phase of the carrier can be used to carry information.
However, in systems where phase information is absent, different modulation techniques must be considered.
Examples of such systems include phase noise limited systems, and noncoherent systems where information is encoded onto the amplitude of the carrier and the envelope of the received signal is detected at the receiver, etc. The latter is prevalent in optical communication systems where the overall cost and complexity is a critical constraint. Such type of noncoherent systems are known as \emph{intensity-modulated direct-detection} (IM/DD) systems and will be the focus of our work. In such systems the information is encoded onto the intensity of the optical carrier, and this intensity must, at all time instances, be nonnegative.
Applications using IM/DD are, for example, wireless optical communications~\cite{Barry1994,Kahn1997,Hranilovic2004a}, and short-haul fiber links including data centers~\cite{Randel2008}.

In the absence of optical amplification, IM/DD systems can be modeled as a conventional AWGN channel whose input is constrained to being nonnegative~\cite[Ch. 5]{Barry1994},~\cite{Kahn1997,Hranilovic2004,Hranilovic2003,Farid2010,Lapidoth2009}.
Since the optical phase cannot be used to carry information, resorting to multilevel pulse amplitude modulation ($M$-PAM) is a natural low-complexity way of extending the widely spread on-off keying (OOK) to improve spectral efficiency.
However, this is different from the conventional PAM since no negative amplitudes can be used~\cite[Eq. (5.8)]{Barry1994}. In~\cite{Cunningham2006}, an IM/DD link analysis using $4$-PAM signaling was performed. In~\cite{Hranilovic2004}, upper and lower bounds on the capacity of $2$-, $4$-, $8$-, and $16$-PAM were derived and in~\cite{Walklin}, the power efficiency of $M$-PAM was shown to be low. The $M$-ary pulse-position modulation ($M$-PPM) are known to be power-efficient; however, they have a poor spectral efficiency~\cite{Hranilovic2005,Kahn1997}.

Since any nonnegative electrical waveform satisfies the above channel constraint, it can be communicated successfully over an IM/DD link. This implies that if the information to be transmitted is firstly modulated on a subcarrier (electrical) using any $M$-level modulation format, it can be transmitted on an IM/DD link after adding a direct current (DC) bias to ensure its nonnegativity, i.e., the subcarrier amplitude and phase which carries the information can be retrieved at the receiver. This concept is known as \emph{subcarrier modulation} (SCM) and was described in the wireless infrared communication context~\cite[Ch. 5]{Barry1994}. 
Therefore, the power efficiency compared to $M$-PAM can be improved since SCM allows the use of power-efficient multilevel modulation formats with IM/DD systems.
In~\cite{Wiberg2009}, the SCM concept is experimentally demonstrated, and in~\cite{Space}, a novel transmitter design for the subcarrier quadrature phase-shift keying (QPSK) and 16-ary quadrature amplitude modulation (16-QAM) is presented.
The DC bias required to ensure the nonnegativity of the electrical waveform does not carry information~\cite[Ch. 5]{Barry1994},~\cite{Wiberg2009,Space}. Therefore, the improved power efficiency can be achieved by allowing the DC bias to vary on a symbol-by-symbol basis and within the symbol interval as in~\cite{Hranilovic1999} and~\cite{You2001}, respectively.
By guaranteeing nonnegativity, the investigation of lattice codes for IM/DD with AWGN became feasible and this is explored in~\cite{Shiu1999,Hranilovic2003}. In~\cite{Hranilovic2003}, a signal space model for optical IM/DD channels is presented, where average and peak optical power are considered as design constraints for constructing lattice-based modulation formats. In addition, constellation shaping to reduce the average optical power has been studied in~\cite{Shiu1999} for the case where no amplification is used, and in~\cite{Mao2008} where optical amplifiers are used.

In this work, we optimize IM/DD modulation formats with and without confining them to a lattice structure. We propose a set of 4- and 8-level single-subcarrier modulation formats which are optimized for average electrical and optical power. These optimization criteria are both relevant, because the average \emph{electrical} power is the standard power measure in digital and wireless communications~\cite[p. 40]{Simon1995} and it helps in assessing the power consumption in optical communications~\cite{Chen1996}, while the average \emph{optical} power is an important figure of merit for skin- and eye-safety measures in wireless optical links~\cite[Ch. 5]{Barry1994},~\cite{Kahn1997,Hranilovic2003} and for quantifying the impact of shot noise in fiber-optical communications~\cite[p. 20]{Cox2002}.

%


%
%

\section{System Model}\label{sec:sysmodel}

\begin{figure}

\psfrag{1}[c][c][1][0]{$u(k)$ }
\psfrag{2}[c][c][1][0]{Modulator }
\psfrag{3}[c][c][1][0]{$x(t) \geq 0$ }
\psfrag{4}[c][c][1][0]{Laser diode }
\psfrag{5}[c][c][1][0]{$z(t)$ }
\psfrag{6}[c][c][1][0]{Optical link }
\psfrag{7}[c][c][1][0]{Photodetector }
\psfrag{8}[c][c][1][0]{$y(t) $ }
\psfrag{9}[c][c][1][0]{Demodulator }
\psfrag{a}[c][c][1][0]{$\hat u(k)$ }
\psfrag{b}[c][c][1][0]{$n(t)$ }
  \centering

  \subfloat{\label{basebandtrans}\resizebox{3in}{!}{\includegraphics{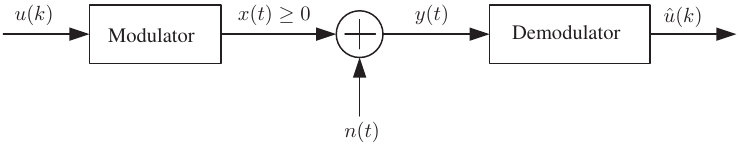}}}
  \hspace{3pt}
  \subfloat[][]{\label{Passbandtrans}
  \psfrag{1}[c][c][1.2][0]{$u(k)$ }
\psfrag{2}[c][c][1.2][0]{Modulator }
\psfrag{3}[c][c][1.2][0]{$x(t) \geq 0$ }
\psfrag{4}[c][c][1.2][0]{Light Source }
\psfrag{5}[c][c][1.2][0]{$z(t)$ }
\psfrag{6}[c][c][1.2][0]{Optical link }
\psfrag{7}[c][c][1.2][0]{Photodetector }
\psfrag{8}[c][c][1.2][0]{$y(t) $ }
\psfrag{9}[c][c][1.2][0]{Demodulator }
\psfrag{a}[c][c][1.2][0]{$\hat u(k)$ }

  \resizebox{3.4in}{!}{\includegraphics{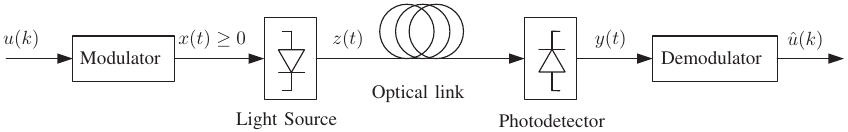}}}

  \caption{(a) Baseband transceiver with constrained-input Gaussian channel. (b) Passband transceiver of IM/DD systems.}
  \label{fig:model}

\end{figure}


The system model under study is depicted in Fig.~\ref{fig:model}(a). It consists of a modulator which maps the data symbols $u(k)$ at instant $k$ to a waveform belonging to the signaling set $S=\{s_0(t),s_1(t), \ldots, s_{M-1}(t)\}$, where $M$ is the size of the signaling set. The generated waveform
\begin{equation}\label{sig:xt}
x(t)=\sum_{l=-\infty}^{\infty} s_{\Gamma[l]}(t-l T_s),
\end{equation}
where $\Gamma[l]$
is an ergodic process uniformly distributed over $\{0,1,\ldots,M-1\}$ and $T_s$ is the symbol period, is constrained to being real and nonnegative.
The received signal can be written as
\begin{equation}\label{basebandreceivedsignal}
    y(t)= x(t) + n(t),
\end{equation}
where $n(t)$ is a zero-mean Gaussian process with double-sided power spectral density $N_0/2$. It should be noted that there exists no nonnegativity constraint on the signal $y(t)$. This is then followed by the demodulation of $y(t)$ which yields $\hat u(k)$, an estimate of $u(k)$. The demodulator is a correlator or matched filter receiver, which minimizes the symbol error rate at a given signal-to-noise ratio (SNR)~\cite[Sect. 4.1]{Simon1995}. This model is different from the conventional AWGN channel by the fact that the input $x(t)$ is constrained to being nonnegative.

The baseband model in Fig.~\ref{fig:model}(a) has been extensively studied in the optical communications context, since it serves as a good model for intensity-modulated direct-detection (IM/DD) systems
~\cite[Ch. 5]{Barry1994},~\cite{Kahn1997,Hranilovic2004,Hranilovic2003,Farid2010,Lapidoth2009}. The passband transceiver for IM/DD systems is depicted in Fig.~\ref{fig:model}(b). In such systems, the electrical nonnegative waveform $x(t)$ directly modulates a light source, such as a laser diode. Therefore, the information is carried on the envelope of the passband signal
$z(t)=\sqrt{2 c x(t)} \cos( 2 \pi f_o t +\theta)$, i.e., the intensity of the optical field, where $c$ represents the electro-optical conversion factor in watts per ampere (W/A)~\cite[pp. 1--33]{Cox2002},~\cite{Westbergh2009},~\cite[pp. 32--67]{Coldren1999}, $f_o$ is the optical carrier frequency, and $\theta$ is a random phase, uniformly distributed in $[0,2\pi)$.
It then propagates through the optical medium depicted as an optical fiber in Fig.~\ref{fig:model}(b), which could be a free-space optical link in other applications. At the receiver, the photodetector detects the power of $z(t)$.
Since the dominant channel impairment in optical IM/DD systems is the thermal noise resulting from the optical-to-electrical conversion~\cite{Mao2008},~\cite[p. 155]{Agrawal2005}, the received electrical signal can be written as
\begin{equation}\label{ytwithslopeefficiency}
    y(t)= r c x(t) + n(t),
\end{equation}
where $r$ is the responsivity of the opto-electrical converter in A/W. Without loss of generality, we set $rc=1$, which yields~(\ref{basebandreceivedsignal}).
%
%

%

%



\section{Signal Space Model}\label{sec:signalspace}
By defining a set of orthonormal basis functions $\phi_k(t)$ for $k=1,2, \ldots,N$ and $N \leq M$ as in~\cite{Hranilovic2003}, each of the signals in $S$ can be represented as
\begin{equation}\label{eachsignal}
    s_i(t)= \sum_{k=1}^{N} s_{i,k} \phi_k(t)
\end{equation}
for $i=0, \ldots, M-1$, where $\mathbf{s}_i=(s_{i,1},s_{i,2}, \ldots, s_{i,N})$ is the vector representation of $s_i(t)$ with respect to the aforementioned basis functions. Therefore, the constellation representing the signaling set $S$ can be written as $\Omega=\{\mathbf{s}_0,\mathbf{s}_1, \ldots, \mathbf{s}_{M-1}\}$.

\subsection{Single-Subcarrier Modulation Formats}\label{sec:scm}
For in-phase and quadrature phase (I/Q) modulation formats to be used on IM/DD channels, a DC bias is required in order for $x(t)$ to be nonnegative. This could be translated geometrically by having a three-dimensional (3d) Euclidean space spanned by the orthonormal basis functions
\begin{eqnarray}\label{basisfunctions}
  \phi_1(t) &=& \sqrt{ \frac{1}{T_s} } ~\rect\left(\frac{t}{T_s}\right) \\ \nonumber
  \phi_2(t) &=& \sqrt{ \frac{2}{T_s} }~\cos{(2\pi f t)} ~\rect\left(\frac{t}{T_s}\right)\\\nonumber
  \phi_3(t) &=& \sqrt{ \frac{2}{T_s} }~\sin{(2\pi f t)}~\rect\left(\frac{t}{T_s}\right),
\end{eqnarray}
where
\begin{eqnarray*}
    \rect(t) =
\begin{cases}
 1 & \mbox{if } 0 \leq t \leq 1 \\
 0 & \mbox{otherwise }
\end{cases}
\end{eqnarray*}
and $f$ is the electrical subcarrier frequency~\cite{Hranilovic2003}. The basis function $\phi_1(t)$ represents the DC bias, where $s_{i,1}$ is chosen for each $i=0,\ldots,M-1$ such that \[\min_{t} s_i(t) \geq 0,\]
which guarantees the nonnegativity of $x(t)$ in~(\ref{sig:xt}). However, $\phi_2(t)$ and $\phi_3(t)$ are the basis functions of the conventional I/Q modulation formats such as $M$-PSK and $M$-QAM.
As in~\cite[pp. 115--116]{Barry1994} and~\cite{Hranilovic2003}, we use $f=1/T_s$, which is the minimum value for which $\phi_1(t)$, $\phi_2(t)$, and $\phi_3(t)$ are orthonormal. In~\cite{Hranilovic2003}, IM/DD modulation formats based on these three basis functions are referred to as raised-QAM, and in~\cite{Westbergha2010} as single-cycle SCM.

\subsection{Performance Measures}
Two important power performance measures can be extracted from the baseband and passband models in Fig.~\ref{fig:model}. The first entity is the average electrical power defined as
\begin{equation*}\label{aveele}
   \bar P_{e}= \lim_{T \to \infty}   \frac{1}{2T} \int_{-T}^{T}    \! x^2(t) \, dt,
\end{equation*}
which for any basis functions can be simplified to
\begin{equation}\label{aveelesig}
   \bar P_{e}=  \frac{\bar{E_s}}{T_s}  =\frac{1}{T_s}~\mathbb{E}[ \|\mathbf{s}_i \|^2] ,
\end{equation}
where $\bar{E_s}$ is the average energy of the constellation and $\mathbb{E}[\cdot]$ is the expected value. This entity is an important figure of merit for assessing the performance of digital and wireless communication systems~\cite[p. 40]{Simon1995}. Therefore, it is relevant for IM/DD systems for compatibility with classical methods and results~\cite{Channels2005,SvalutoMoreolo2010}. In addition, it helps in quantifying the impact of relative intensity noise (RIN) in fiber-optical links~\cite[pp. 1--33]{Cox2002}, and in assessing the power consumption of optical systems~\cite{Chen1996}. In~\cite{Karout2010}, $\bar P_{e}$ was used as a performance measure for comparing different intensity modulation formats.

The second measure is the average optical power $\bar P_o$, which has been studied in~\cite{Barry1994,Kahn1997,Hranilovic2004,Hranilovic2003,Farid2010} for the wireless optical channel. Limitations are set on $\bar P_o$ for skin- and eye-safety standards to be met. In fiber-optic communications, this entity is used to quantify the impact of shot noise on the performance~\cite[p. 20]{Cox2002}. It is defined as
 \begin{equation*}\label{opticalpower}
    \bar P_{o}=  \lim_{T\to \infty}\frac{1}{2T} \int_{-T}^{T} \! z^2(t) \, dt= \lim_{T \to \infty} \frac{c}{2T} \int_{-T}^{T} \!  x(t) \, dt.
  \end{equation*}
This measure depends solely on the DC bias required to make the signals nonnegative and can be represented in terms of the symbol period and the constellation geometry as~\cite{Hranilovic2004,Hranilovic2003}
\begin{equation}\label{opticalpowersig}
    \bar P_{o}=  \frac{c}{\sqrt{T_s}}~\mathbb{E} [s_{i,1}].
      \end{equation}

In order to have a fair comparison between the different modulation formats, the spectral efficiency defined as
$
\eta={R_b/W}
$ (bits/s/Hz)
should be taken into account, where $R_b=R_s \log_2M$ is the bit rate, $R_s=1/T_s$ is the symbol rate, and $W$ is the baseband bandwidth defined as the first null in the spectrum of $x(t)$. The term baseband bandwidth is due to the fact that the baseband model in Fig.~\ref{fig:model}(a) hides the high frequency nature of the optical carrier~\cite[pp. 109--116]{Barry1994}. At the same symbol rate, modulation formats such as OOK and $M$-PAM have $W=R_s$, whereas the modulation formats belonging to the single-subcarrier family in Sec.~\ref{sec:scm} occupies $W=2 R_s$; this is due to the intermediate step of modulating the information onto an electrical subcarrier before modulating the optical carrier~\cite[Ch. 5]{Barry1994},~\cite{Karout2010}.


\begin{figure*}

\begin{tabular}{cccc}
\includegraphics[width=0.23\textwidth]{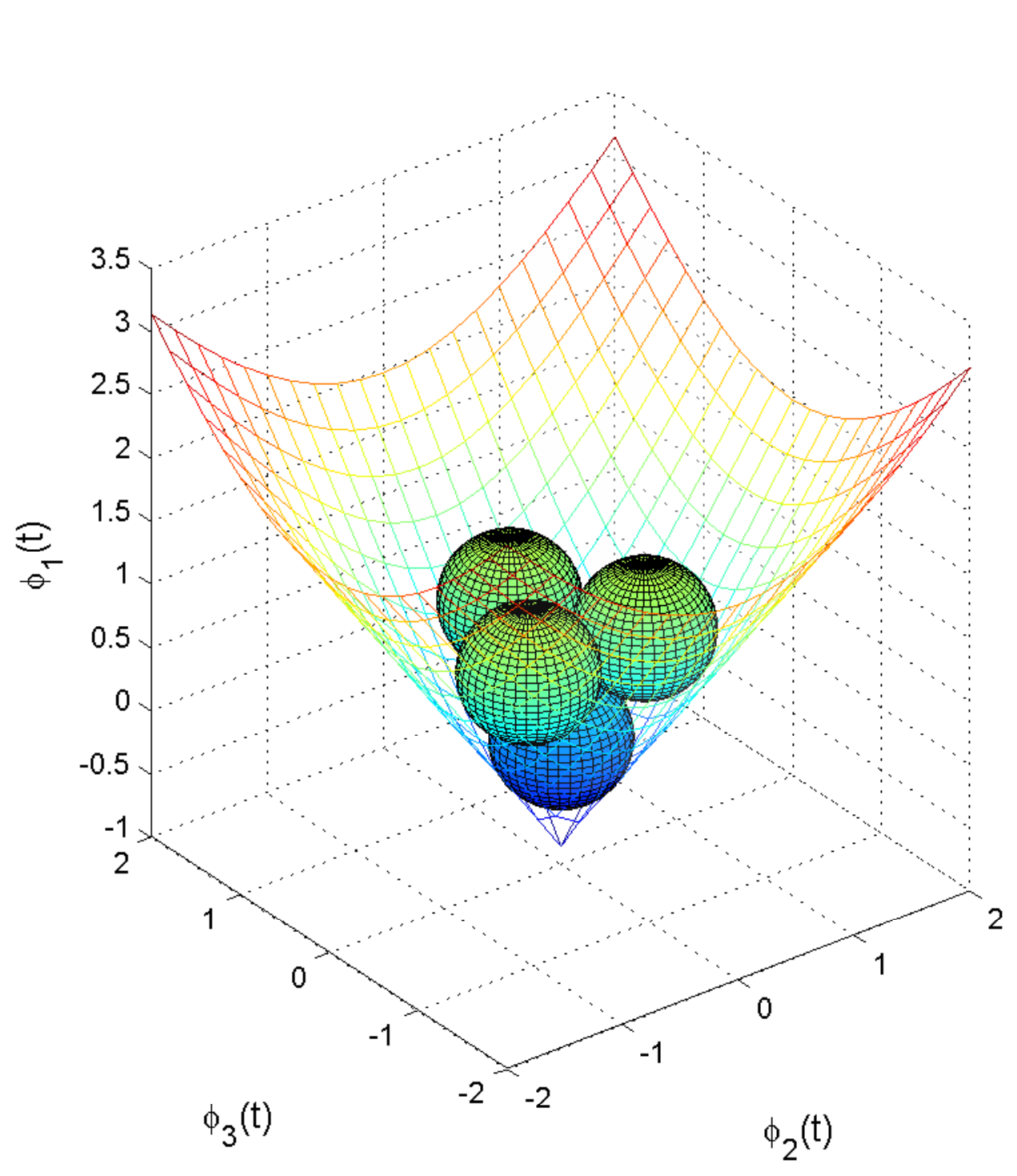}  &
\includegraphics[width=0.23\textwidth]{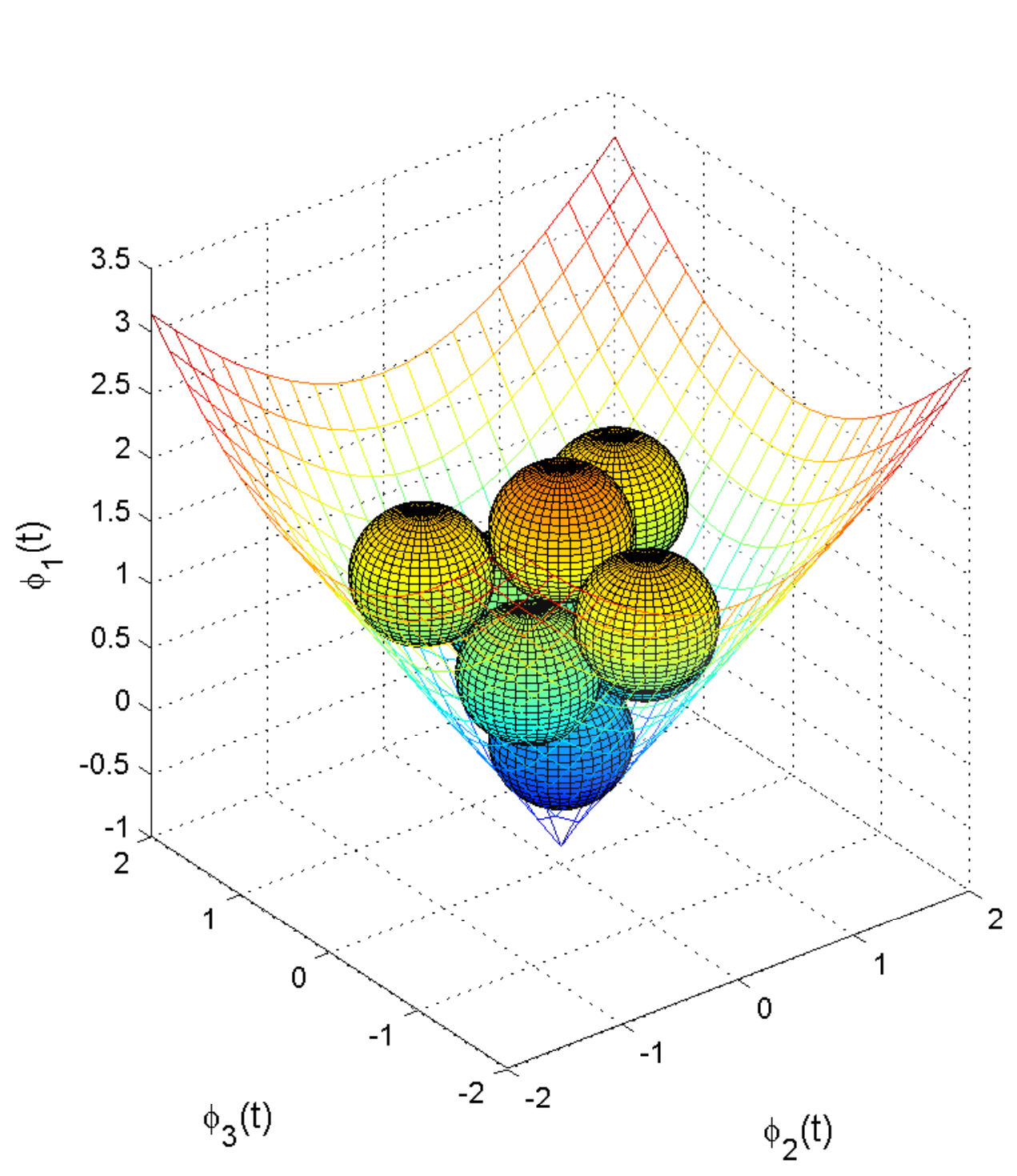} &
\includegraphics[width=0.23\textwidth]{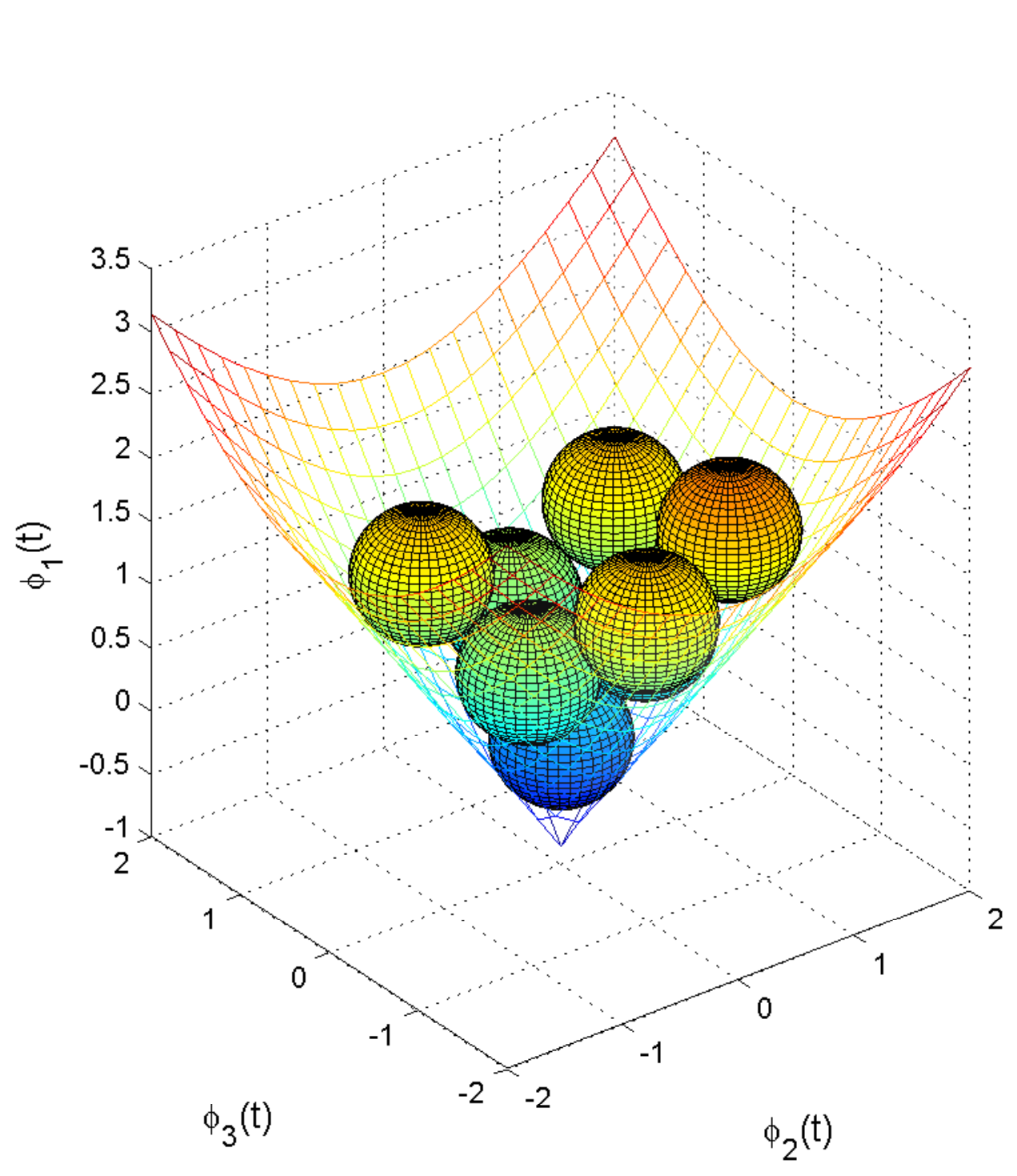}&
\includegraphics[width=0.23\textwidth]{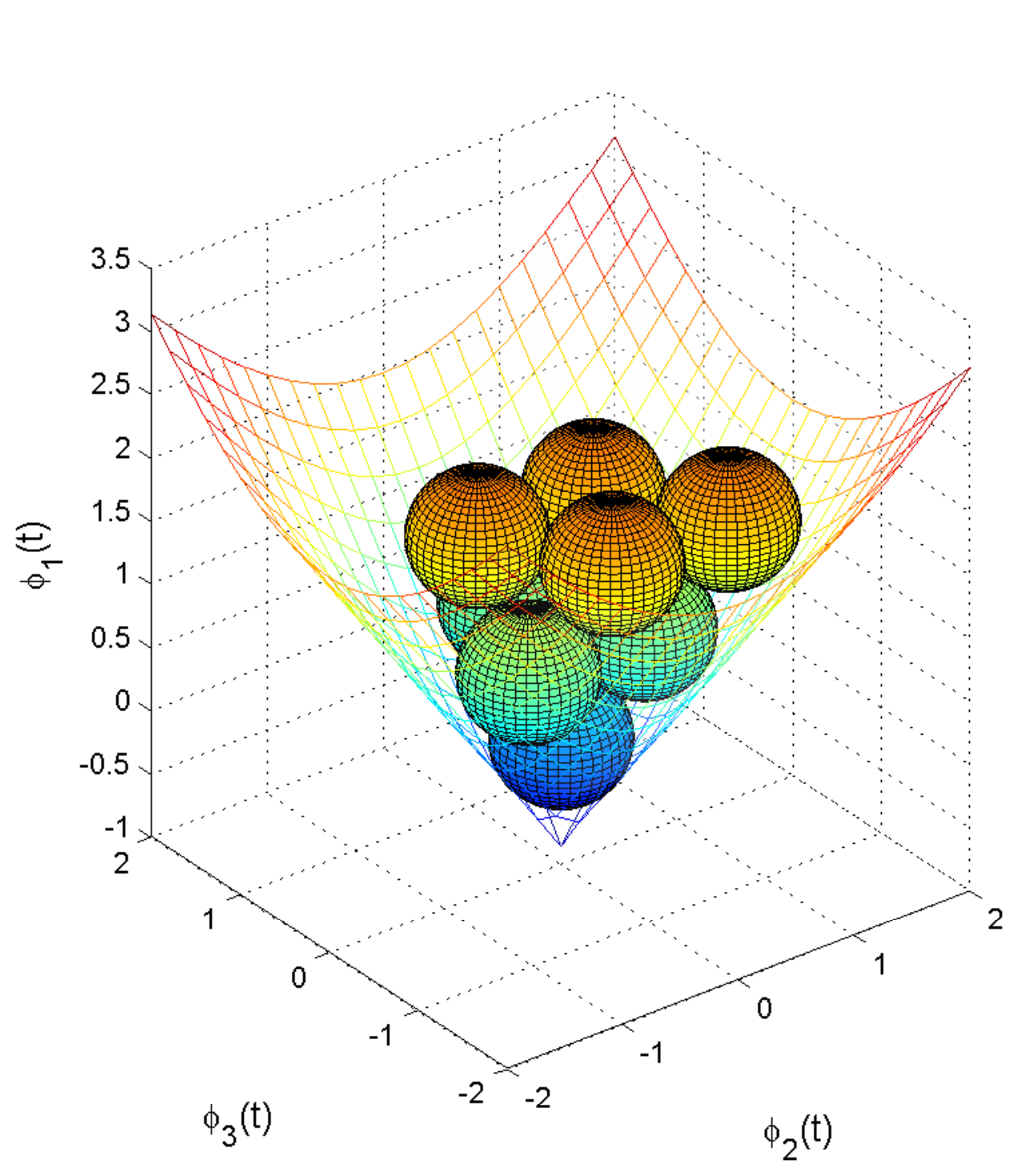}
\end{tabular}
\caption{(left to right):~$\mathscr{C}_{4}$ or $\mathscr{L}_4$, $\mathscr{C}_{\bar P_e,8}$, $\mathscr{C}_{\bar P_o,8}$, and $\mathscr{L}_8$.}
\label{4pointscone}
\label{8pointscone}

\end{figure*}

\section{Constellation Optimization}\label{sec:const}

For designing constellations, the admissible region
defined in~\cite[Eq. (10)]{Hranilovic2003} as the set of all signals satisfying the nonnegativity constraint has to be taken into account. Given the 3d-Euclidean space defined for SCM in Sec.~\ref{sec:scm}, the admissible region can be written as
\begin{equation}\label{adm:scm}
    \Upsilon= \{\mathbf{s} \in \mathbb{R}^3 : s_{i,1}^2 \geq 2(s_{i,2}^2 +s_{i,3}^2)\},
\end{equation}
which  is a 3d-cone with apex angle of $\cos^{-1}(1/3)=70.528^{\circ}$ pointing in the dimension spanned by $\phi_1(t)$. 

As done before for the conventional AWGN channel~\cite{Foschini1974,Porath2003,Sloane1995,Graham1990,Agrell2009}, our approach of finding the best constellations can be formulated as a sphere-packing problem with the objective of minimizing a cost function depending on the constraints that might be present in the system model shown in Fig.~\ref{fig:model}. Thus, the optimization problem, for given constants $M$ and $d_{\text{min}}$, can be written as
\begin{eqnarray}\label{obj}
\label{minline} \text{Minimize~} & & \xi(\Omega)  \\
\label{minline1} \text{Subject to}& & |\Omega| =M\\
\label{cons2}   & & \Omega \subset \Upsilon\\
\label{cons1}   & &  d(\Omega) = d_{\text{min}},
\end{eqnarray}
where \[d(\Omega)=\min_{  \underset{i \ne j }{\mathbf{s}_i ,\mathbf{s}_j \in \Omega}} \| \mathbf{s}_i -\mathbf{s}_j \|.\]

Choosing the objective function as $\xi(\Omega)=\mathbb{E}[ \|\mathbf{s}_i \|^2]$ results in $\Omega=\mathscr{C}_{\bar P_e,M}$, i.e., a constellation optimized for average electrical power, and $\xi(\Omega)=\mathbb{E} [s_{i,1}]$ results in $\Omega=\mathscr{C}_{\bar P_o,M}$, i.e., a constellation optimized for average optical power. The constraint in~(\ref{cons2}) guarantees that the signals belong to the admissible region $\Upsilon$, therefore satisfying the nonnegativity criterion of the channel. 
The minimum distance $d_{\text{min}}$ serves as a good measure of error probability performance in the presence of AWGN at high SNR.
Although this optimization problem is well formulated mathematically, it is rather difficult to obtain an analytical solution. Therefore, we resorted to numerical optimization techniques as in~\cite{Foschini1974,Porath2003,Sloane1995,Graham1990,Agrell2009} to find the best constellations. The optimization problem is nonconvex; therefore, a local solution does not imply that it is global.

A special case of this optimization problem, which might not guarantee the optimal solution, is to confine the possible constellations to have a regular structure such as that of a lattice, denoted by $\Lambda$. In this case, the above optimization problem can be reformulated by replacing~(\ref{cons2}) with  $ \Omega \subset \Upsilon \cap \Lambda$, and dropping~(\ref{cons1}) since it is directly inferred by~(\ref{cons2}). By using the face-centered cubic lattice ($A_3$), which provides the densest packing for the 3d-Euclidean space~\cite[p.~\rmnum{16}  ]{Conway1999}, we obtain $\mathscr{L}_{\bar P_e,M}$, and $\mathscr{L}_{\bar P_o,M}$, the constellations optimized for average electrical and optical power, respectively.

\subsection{Optimized Constellations}
Below is the description of the obtained constellations whose coordinates are included in App.~\ref{bestconstapp}. 
We conjecture that all of them are optimal solutions of~(\ref{minline})--(\ref{cons1}).


\subsubsection{$4$-level Constellations}
Fig.~\ref{4pointscone} depicts the 4-level constellation which provides the lowest $\bar P_e$ and $\bar P_o$ while satisfying the optimization constraints. The geometry of this constellation is a regular tetrahedron where all the spheres, or the constellation points lying at the vertices of this regular tetrahedron, are equidistant from each other and normalized to unit $d_{\text{min}}$. 
This constellation is also the result of $\Upsilon \cap A_3$. Since the obtained constellation is optimized for both $\bar P_e$ and $\bar P_o$, we will refer to it as $\mathscr{C}_4$ or $\mathscr{L}_4$. 
It is a remarkable fact that the vertex angle of the tetrahedron, defined as the apex angle of the circumscribed cone, is exactly $\cos^{-1}(1/3)$, which is equal to the apex angle of the admissible region $\Upsilon$. Thus, $\mathscr{C}_4$ fits $\Upsilon$ snugly, in the sense that all constellation points, regarded as unit-diameter spheres, touch each other as well as the boundary of $\Upsilon$, which, as we shall see in the next section, makes the modulation format very power-efficient.
This modulation consists of a zero level signal and a biased ternary PSK. 
In~\cite{Karout2010}, a power-efficient modulation format called on-off phase-shift keying (OOPSK) was presented. It turns out that it has the same geometry as $\mathscr{C}_4$.
%
%
Other hybrids between amplitude-shift keying and PSK have been studied in~\cite{Essiambre2010} and \cite{Gursoy2009}; however, such modulation formats do not satisfy the nonnegativity constraint of IM/DD channels.


\subsubsection{$8$-level Constellations}
Fig.~\ref{8pointscone} also shows the 8-level constellations $\mathscr{C}_{\bar P_e,8}$ and $\mathscr{C}_{\bar P_o,8}$. None of these constellations are lattice-based, but all of them contain $\mathscr{C}_4$ as the lowest four spheres. The highly symmetric and compact constellation $\mathscr{C}_{\bar P_e,8}$ consists of four central spheres arranged in a tetrahedron and four additional spheres, each touching three spheres in the central tetrahedron. Surprisingly, seven of the eight spheres touch the conical boundary of $\Upsilon$. This modulation is a hybrid between 2-PAM and two ternary PSK which are DC biased differently. The constellation $\mathscr{C}_{\bar P_o,8}$ is the same as $\mathscr{C}_{\bar P_e,8}$ but with one sphere moved. On the other hand, when confining the set of points to a lattice structure, the resulting constellations which provide the lowest $\bar P_e$ and $\bar P_o$ are the same, $\mathscr{L}_{8}=\mathscr{L}_{\bar P_e,8}=\mathscr{L}_{\bar P_o,8}$, and is depicted in Fig.~\ref{8pointscone}.  This constellation also contains $\mathscr{C}_4$ as the lowest four spheres.

\subsection{Previously Known Constellations}
Our investigation encompasses other previously known formats which are presented after being normalized to unit $d_{\text{min}}$. At spectral efficiency $\eta=1$ bits/s/Hz, OOK defined as $\{(0),(1)\}$ in terms of $\phi_1(t)$ will be compared with $\mathscr{C}_4$, and with subcarrier QPSK defined as $\{(1,\pm 1/2,\pm 1/2)\}$ in terms of the basis functions defined in Sec.~\ref{sec:scm}.

At a spectral efficiency $\eta=3/2$ bits/s/Hz, $\mathscr{C}_{\bar P_e,8}$, $\mathscr{C}_{\bar P_o,8}$, and $\mathscr{L}_{8}$ will be compared with subcarrier 8-PSK defined as
$(1/\sin(\pi/8))\{(1/ \sqrt{2}  , \cos( \pi i/4)/2 ,  \sin( \pi i/4) / 2  )\}$
for $i=0,\ldots,7$,
star-shaped 8-QAM~\cite{Essiambre2010} with a constant bias defined as
$\{ ( (1+\sqrt{3})/\sqrt{2},\pm 1/2 ,\pm 1/2  ), ((1+\sqrt{3})/\sqrt{2},0,\pm (1+\sqrt{3})/2),((1+\sqrt{3})/\sqrt{2},\pm (1+\sqrt{3})/2,0) \}$. We also include in our analysis a star-shaped 8-QAM denoted as $\breve 8$-QAM in which the DC bias is allowed to vary from symbol to symbol, thus carrying information, and is defined as
$\{ ( 1,\pm 1/2 ,\pm 1/2  ), ((1+\sqrt{3})/\sqrt{2},0,\pm (1+\sqrt{3})/2),((1+\sqrt{3})/\sqrt{2},\pm (1+\sqrt{3})/2,0) \}$.

\begin{figure}
\centering
\footnotesize
\psfrag{aaaaaaaa1}[c][c][0.75][0]{$\mathrm{OOK}$}
 \psfrag{aaaaaaaa2}[c][c][0.75][0]{QPSK }
\psfrag{aaaaaaaa3}[c][c][0.9][0]{$\mathscr{C}_4$ }
\psfrag{aaaaaaaa4}[c][c][0.75][0]{8-QAM}
 \psfrag{aaaaaaaa5}[c][c][0.75][0]{$\mathscr{L}_8$}
\psfrag{aaaaaaaa6}[c][c][0.9][0]{$\mathscr{C}_{\bar P_e,8}$}
\psfrag{aaaaaaaa7}[c][c][0.9][0]{$\mathscr{C}_{\bar P_o,8}$}
\psfrag{aaaaaaaa8}[c][c][0.9][0]{$\mathscr{C}_{\hat P_o,8}$}

\psfrag{xxxxxxxx4}[c][c][0.75][0]{$\breve 8$-QAM}
\psfrag{xxxxxxxx5}[c][c][0.75][0]{8-PSK}

\psfrag{SER}[c][c][1][0]{$P_s$}
\psfrag{SNREbN0dB}[t][c][1][0]{$\gamma_{\bar E_b}~\text{[dB]}$}
\psfrag{TbavePosquaredNodB}[t][c][1][0]{$\gamma_{\bar P_o}~\text{[dB]}$}
\psfrag{TbpeakPosquaredNodB}[t][c][1][0]{$\gamma_{\hat P_o}~\text{[dB]}$}
\begin{tabular}{c}
\includegraphics[scale=0.9]{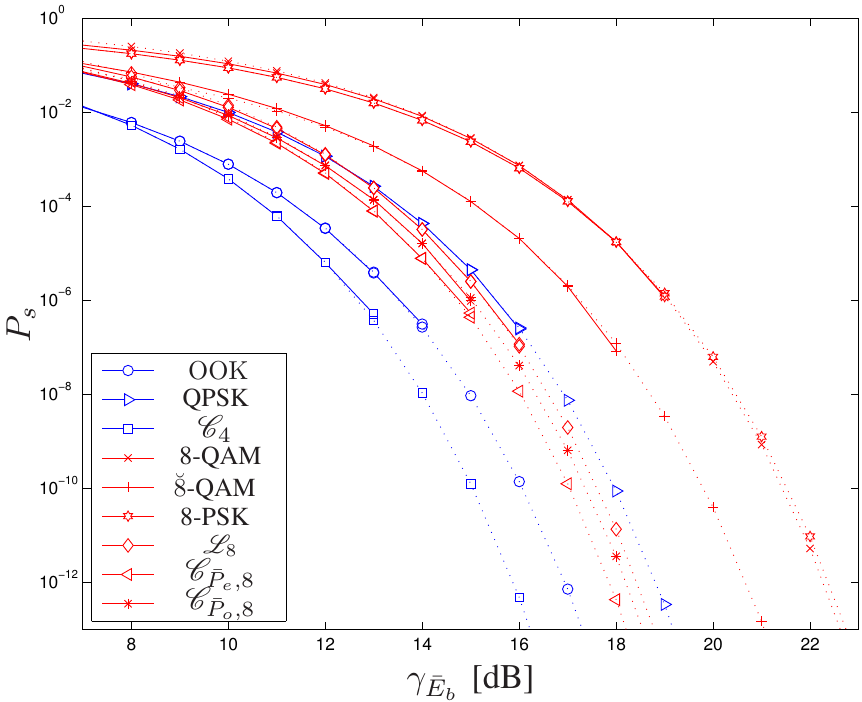}    \\ \\
\includegraphics[scale=0.9]{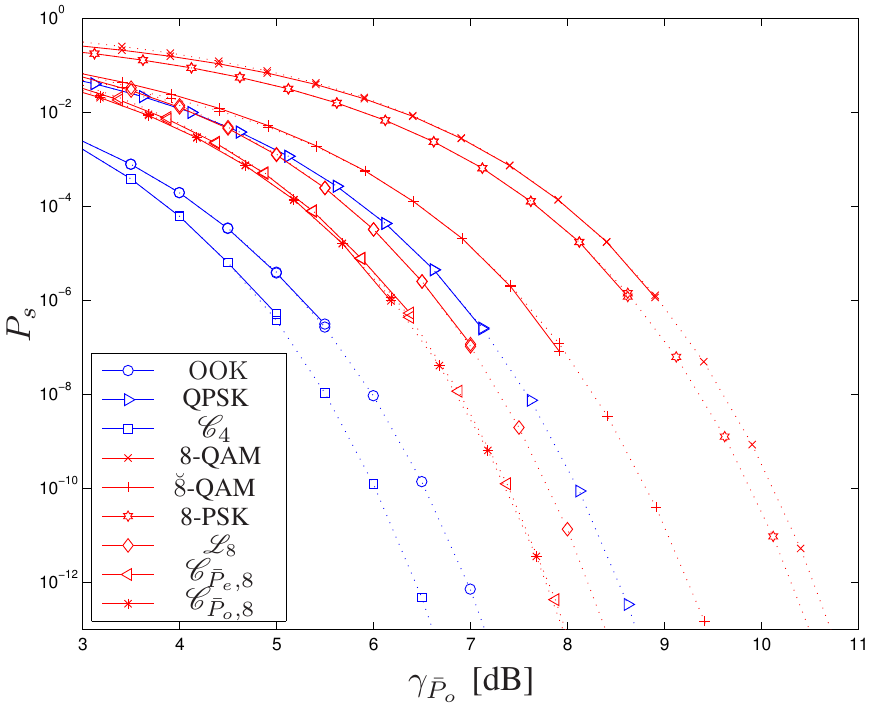} \\ \\
\end{tabular}
\caption{Simulated (solid) and theoretical (dotted) SER for the various modulation formats  vs. $\gamma_{\bar E_b}$ (top), $\gamma_{\bar P_o}$ (bottom).}
\label{serebno}
\end{figure}


\section{Performance Analysis}\label{sec:perf}

The symbol error rate (SER) performance of the different modulation schemes will be assessed. The standard union bound found in~\cite[Eq. (4.81)]{Simon1995} is used to approximate the theoretical SER. This union bound can be approximated as
\begin{equation}\label{bound}
    P_{  s   } \approx \frac{2 K }{M }~Q\left(   \sqrt{\frac{ d_{\text{min}}^2}{2 N_0}}  \right),
\end{equation}
where $K$ is the number of distinct signal pairs $(s_i(t),s_j(t))$ with $i<j$ for which $\int (s_i(t)-s_j(t))^2 dt = d_{\text{min}}^2$. This approximation is tight at high SNR.

%

Fig.~\ref{serebno} (top) shows the simulated and theoretical SER of the studied modulation formats vs. SNR defined as
\begin{equation}\label{snreb}
    \gamma_{\bar E_b} =10 \log_{10}\frac{\bar E_b}{N_0}.
\end{equation}
Apparently, all the modulation formats which are optimized for $\bar P_e$ outperforms the other formats at the same spectral efficiency.
For spectral efficiency $\eta=1$, $\mathscr{C}_4$ has a
0.86 dB average electrical power gain over OOK 
and 2.87 dB gain over QPSK to achieve $P_s= 10^{-6}$.
For $\eta=3/2$,
$\mathscr{C}_{\bar P_e,8}$ has a
0.3 dB gain over $\mathscr{C}_{\bar P_o,8}$,
0.58 dB gain over $\mathscr{L}_{8}$,
2.55 dB gain over $\breve 8$-QAM,
4.35 dB gain over 8-QAM,
and 4.39 dB gain over 8-PSK to achieve $P_s= 10^{-6}$.
The modulation formats optimized for $\bar P_e$ and $\bar P_o$ are very close in performance to $\mathscr{L}_{8}$, the lattice-based modulation format.

In order to facilitate the comparison of modulation formats in terms of their average optical power requirements, we define the optical SNR as
\begin{equation}\label{snravePo}
    \gamma_{\bar P_o} = 10 \log_{10}\frac{  \bar P_o}{c\sqrt{R_b N_0}}
\end{equation}
in a similar fashion as in~\cite[Eq. (5)]{Kahn1997} . Fig.~\ref{serebno} (bottom) shows the SER plotted vs. $\gamma_{\bar P_o}$. Quite obviously, the modulation formats optimized for $\bar P_o$ perform better than the rest.
For $\eta=1$, $\mathscr{C}_4$ has a
0.43 dB average optical power gain over OOK, and a
2.06 dB gain over QPSK to achieve an SER of $10^{-6}$.
For $\eta=3/2$, $\mathscr{C}_{\bar P_o,8}$ has a
0.04 dB gain over $\mathscr{C}_{\bar P_e,8}$,
0.46 dB gain over $\mathscr{L}_8$,
1.35 dB gain over $\breve 8$-QAM,
2.48 dB gain over 8-PSK,
and a 2.75 dB gain over 8-QAM to achieve a $P_s= 10^{-6}$.
Besides the very close performance of $\mathscr{C}_{\bar P_o,8}$ and $\mathscr{C}_{\bar P_e,8}$, it is clear that they together with the lattice-based modulation perform better than the other modulation formats under study in terms of average optical power performance.

\section{Conclusion}\label{sec:conclusion}

By relaxing the constraint on the set of points used to design modulation formats for IM/DD channels, we were able to design 4- and 8-level modulation formats which are more power-efficient than known ones. For the 4-level modulation formats, the most power-efficient modulation in terms of average electrical and optical power happens to have a lattice structure even though the number of constellation points is small. This constellation is also a subset of all the obtained higher level constellations. As for the 8-level constellations, power-efficient schemes are obtained by not confining the set of constellation points to a lattice structure. However, this comes at the price of losing the geometric regularity found in the lattice structure. We conjecture that the new modulation formats are optimal for their size and optimization criteria over IM/DD channels.

\appendices

\section{Obtained Constellations}\label{bestconstapp}
Constellations are normalized to unit $d_{\text{min}}$. 
 \vspace{-4mm}

\begin{align*}
 \mathscr{C}_{\bar P_e,4}= & \mathscr{C}_{\bar P_o,4}=\mathscr{L}_{\bar P_e,4}=\mathscr{L}_{\bar P_o,4}=\{(0 , 0, 0),\\
&(\sqrt{2/3} , 0 , 1/\sqrt{3}),
(\sqrt{2/3},  \pm 1/2 , -\sqrt{3}/6)\}.\\ 
 \mathscr{C}_{\bar P_e,8}= & \mathscr{C}_4 \cup \{
 ((5/3)\sqrt{2/3}  ,0,  -5/(3\sqrt{3})),\\
&((5/3)\sqrt{2/3} ,\pm 5/6, 5/(6\sqrt{3})),
(2 \sqrt{2/3}   ,                0  ,                 0)
\}.\\ 
\mathscr{C}_{\bar P_o,8}= & \mathscr{C}_4 \cup \{
((5/3)\sqrt{2/3},  0, -5/(3\sqrt{3})),\\
&((5/3)\sqrt{2/3},   \pm 5/6 ,  5/(6\sqrt{3})),\\
&(1.6293   ,0.9236,  -0.6886)
\}.\\ 
 \mathscr{L}_{\bar P_e,8}= & \mathscr{L}_{\bar P_o,8}=\mathscr{C}_4 \cup \{
 (2\sqrt{2/3}, \pm1/2, \sqrt{3}/6),\\
&(2\sqrt{2/3},         0, -1/\sqrt{3}),
(        2\sqrt{2/3}, 1, -1/\sqrt{3})
\}.
\end{align*}

\section*{Acknowledgment}

The authors would like to acknowledge the SSF funding under grant RE07-0026 and LINDO Systems for the free license to use their numerical optimization software.

\ifCLASSOPTIONcaptionsoff
  \newpage
\fi

\begin{IEEEbiography}{Michael Shell}
Biography text here.
\end{IEEEbiography}

\begin{IEEEbiographynophoto}{John Doe}
Biography text here.
\end{IEEEbiographynophoto}


\begin{IEEEbiographynophoto}{Jane Doe}
Biography text here.
\end{IEEEbiographynophoto}




\end{document}